\documentclass[format=acmsmall]{acmart}

\usepackage[english]{babel}
\usepackage[utf8]{inputenc}
\usepackage[colorinlistoftodos]{todonotes}
\usepackage{xcolor}
\usepackage{xurl}
\usepackage{soul}

\setcopyright{rightsretained}
\copyrightyear{2023}
\acmYear{2023}
\acmDOI{XXXXXXX.XXXXXXX}

\begin{document}

\title[A Review of Deep Reinforcement Learning in Serverless Computing]{A Review of Deep Reinforcement Learning in Serverless Computing: Function Scheduling and Resource Auto-Scaling
}


\author{Amjad Yousef Majid}
\affiliation{%
  \institution{Martel Innovate}
  \streetaddress{Keizersgracht 482}
  \city{Amsterdam}
  \country{the Netherlands}}
\email{amjad.majid@martel-innovate.com}

\author{Eduard Marin}
\affiliation{%
  \institution{Telefonica}
  \streetaddress{1 Th{\o}rv{\"a}ld Circle}
  \city{Madrid}
  \country{Spain}}
\email{eduard.marinfabregas@telefonica.com}

\renewcommand{\shortauthors}{A.Y. Majid, et al.}

\begin{abstract}
In the rapidly evolving field of serverless computing, efficient function scheduling and resource scaling are critical for optimizing performance and cost. 
This paper presents a comprehensive review of the application of Deep Reinforcement Learning (DRL) techniques in these areas. 
We begin by providing an overview of serverless computing, highlighting its benefits and challenges, with a particular focus on function scheduling and resource scaling.
We then delve into the principles of deep reinforcement learning (DRL) and its potential for addressing these challenges. 
A systematic review of recent studies applying DRL to serverless computing is presented, covering various algorithms, models, and performances. 
Our analysis reveals that DRL, with its ability to learn and adapt from an environment, shows promising results in improving the efficiency of function scheduling and resource scaling in serverless computing. 
However, several challenges remain, including the need for more realistic simulation environments, handling of cold starts, and the trade-off between learning time and scheduling performance. 
We conclude by discussing potential future directions for this research area, emphasizing the need for more robust DRL models, better benchmarking methods, and the exploration of multi-agent reinforcement learning for more complex serverless architectures. 
This review serves as a valuable resource for researchers and practitioners aiming to understand and advance the application of DRL in serverless computing.
\end{abstract}





\maketitle

\section{Introduction}

The advent of serverless computing has revolutionized the way applications are developed and deployed in the cloud. By abstracting away the underlying infrastructure, serverless computing allows developers to focus on writing applications' code, while the cloud provider manages the execution environment, including server management, capacity planning, and scaling. This paradigm shift, however, introduces new challenges in terms of function scheduling and resource scaling, which are critical for optimizing performance and cost in serverless environments.

Function scheduling in serverless computing involves deciding when and where to execute a function, considering factors such as resource availability, current workload, performance security requirements, or privacy regulations \cite{ghobaei2023scheduling}. 
Resource scaling, on the other hand, refers to the dynamic allocation and deallocation of resources based on demand~\cite{mampage2022holistic}. 
Both of these tasks are complex and dynamic problems that require intelligent and adaptive solutions.

Traditional ML algorithms, while powerful, were designed for more static datasets and predetermined variables. In the dynamic environment of serverless computing, where demands and patterns can change rapidly, these algorithms might not adapt quickly enough or may require extensive retraining. Additionally, they typically do not handle the real-time, adaptive decision-making that's essential for optimizing serverless deployments, given the importance of instantaneous feedback. For example, 
Smart Spread uses supervised ML for serverless function scheduling. Consequently, functions have to be profiled before deployment which is considered a critical drawback~\cite{mahmoudi2019optimizing}. 

Deep Reinforcement Learning (DRL), a subfield of artificial intelligence that combines deep learning and reinforcement learning, has shown promise in addressing these challenges \cite{yao2023performance, qiu2022simppo, mampage2023deep}. 
DRL models learn to make decisions by interacting with an environment, receiving feedback in the form of rewards or penalties, and adjusting their strategies to maximize the cumulative reward \cite{majid2023deep}. This ability to learn and adapt makes DRL particularly suitable for the dynamic and uncertain environments of serverless computing.

There are many review and survey papers that target a variety of serverless computing aspects~\cite{cassel2022serverless,wen2022literature,li2022serverless,shafiei2022serverless,hassan2021survey, jawaddi2023autoscaling}. The use of DRL for cloud orchestration---a process that coordinates and automates the management and control of the computer systems, middleware, and services within a cloud environment to ensure they operate in harmony. Serverless function scheduling is intrinsically linked to cloud orchestration as it requires efficiently coordinating where and when serverless functions run---and resource management has also been surveyed and reviewed~\cite{zhong2022machine,duc2019machine,kar2023offloading}. However, there is no paper that is dedicated to reviewing or surveying that work on the use of (D)RL for optimizing 
both function scheduling and resource scaling in
serverless environments. This paper aims to fill this gap by providing a systematic review of recent studies that apply DRL to serverless computing. We cover various DRL algorithms and models, evaluate their performance metrics, and discuss the challenges and future directions in this research area. This review serves as a valuable resource for researchers and practitioners aiming to understand and advance the application of DRL in serverless computing.

\section{Background}

\subsection{Serverless Computing}
Serverless computing, also known as Function as a Service (FaaS), is a cloud computing model where the cloud provider dynamically manages the allocation and provisioning of servers allowing users to write and deploy code without worrying about the underlying infrastructure~\cite{jonas2019cloud}. Key benefits of serverless computing include reduced operational concerns, automatic scaling, and cost-effectiveness. However, it also presents unique challenges, such as statelessness, cold start latency, and the need for efficient function scheduling and resource auto-scaling in (near) real time~\cite{li2022serverless}.


\textit{Serverless Scheduling} is about how tasks are assigned and managed within a serverless architecture. In a serverless environment, the cloud provider manages the server infrastructure, and tasks or functions are executed in stateless compute containers that are event-triggered and fully managed by the cloud provider. Serverless scheduling involves deciding when and where these tasks or functions are run. It's about efficiently distributing tasks among the available resources to optimize for factors like performance, cost, and completion time while taking security and privacy into consideration. 

\textit{Resource Auto-scaling} is the dynamic capability of cloud infrastructures to automatically tailor computational resources according to the demand~\cite{kriushanth2013auto}. It's not a monolithic approach. While serverless computing often provides a pre-defined set of computational resources for functions, such as a fixed memory size, advanced platforms now allow diverse computational requirements to cater to varying workloads~\cite{mampage2022holistic}. This complexity highlights the challenge and importance of effective auto-scaling. Essentially, auto-scaling dynamically adjusts resources like servers, memory, or storage based on workload requirements. Overestimating needs lead to under-utilized resources, resulting in increased costs. Conversely, underestimating can compromise the application's performance or even lead to its failure. Auto-scaling can be predictive, where resources are allocated based on a schedule, or reactive, adjusting to real-time workload changes~\cite{poppe2022moneyball}. The potential of Deep Reinforcement Learning (DRL) extends to enhancing both predictive and reactive scaling strategies. To encapsulate, while both processes focus on efficient resource utilization in a cloud setting, resource auto-scaling modifies the volume of resources, whereas serverless scheduling oversees the task executions within those allocated resources.

\textit{Serverless versus Microservices}
Although autoscaling in serverless computing and microservices are driven by the same objective of adjusting resources based on workload, they differ fundamentally in their mechanisms~\cite{castro2019rise,jawaddi2022review}. In serverless computing, resources are automatically allocated for each function or task as it's triggered, with execution occurring in isolated environments. Scaling is virtually limitless, and users are charged solely for the actual compute time. Contrarily, microservices autoscaling necessitates a hands-on approach, adjusting the number of individual microservice instances as required. This involves active performance monitoring, orchestration tools, and an understanding of the infrastructure, leading to charges based on reserved resources irrespective of their full utilization~\cite{toffetti2017self}.

Function scheduling in serverless platforms is largely abstracted, with events triggering functions and the platform ensuring seamless execution. Resource allocation in this context is handled internally, with automatic assignment of required compute, memory, and storage~\cite{kaffes2022hermod}. In contrast, microservices architectures require more direct engagement: scheduling determines which instances handle requests, often involving tools like load balancers, and resource allocation decisions are more explicit, demanding careful capacity planning~\cite{toffetti2017self}. While serverless autoscaling simplifies infrastructure management, microservices autoscaling offers increased control, albeit with greater complexity.

\subsection{Deep Reinforcement Learning}
    \begin{figure}
        \centering
        \includegraphics[width=.7\columnwidth]{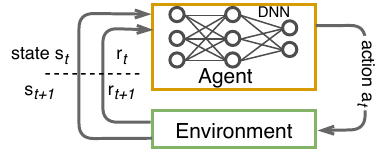}
        \caption{Deep Reinforcement Learning Overview} 
        \label{fig:DRL_fig}
    \end{figure}

Deep Reinforcement Learning (DRL) is a subfield of artificial intelligence that combines the strengths of deep learning and reinforcement learning. 
Deep learning is a type of machine learning that uses neural networks with many layers to model and understand complex patterns in data. 
Reinforcement learning, on the other hand, is a type of machine learning where an agent learns to make decisions by interacting with an environment and receiving feedback in the form of rewards or penalties. 
The combination of these two fields in DRL allows for the creation of models that can handle high-dimensional inputs, deal with sequential data, and learn from interaction with an environment. 
This interaction is often modeled as a Markov Decision Process (MDP;  Figure \ref{fig:DRL_fig}). An MDP includes:  
(i) Agent: The entity that makes decisions and takes actions; 
(ii) Environment: The context in which the agent operates and interacts.
(iii) State: The current situation or condition of the environment;
(iv) Action: A choice made by the agent that affects the state of the environment;
(v) Reward: Feedback from the environment that indicates the success of an action. Rewards can be positive (for successful actions) or negative (for unsuccessful actions);
(vi) Policy: The strategy that the agent uses to decide which action to take in a given state. In DRL, the policy is typically represented by a deep neural network.

The objective of a DRL agent is to maximize the total reward (or return) over a series of interactions with an environment.
This is facilitated by value functions, namely the state-value function and the action-value function (or Q-function), which provide the expected return of being in a particular state or performing a specific action and following a specific policy thereafter~\cite{sutton2018reinforcement}. 

Several types of RL algorithms have been developed. They are largely categorized under three main categories: 
\begin{itemize}
    \item Value-based RL algorithms optimize value functions to select the most rewarding actions. Examples of such algorithms are SARSA~\cite{SARSA}, Q-learning~\cite{sutton2018reinforcement} and DQN~\cite{Mnih2015} which combines Q-learning with deep neural networks (DNNs). 

    \item Policy-based RL algorithms optimize the policy directly, outputting a probability distribution over the action space given a state. These algorithms are particularly useful for learning stochastic policies and dealing with high-dimensional or continuous action spaces, despite potentially being sample inefficient. An example of this is the REINFORCE algorithm~\cite{WilliamsREINFORCE1992}.

    \item Actor-critic algorithms combine policy and value function learning to reduce variance and speed up learning. These algorithms use a policy "actor" and a value estimator "critic", both parameterized by DNNs. Prominent examples include the Advantage Actor-Critic (A2C)~\cite{clemente2017efficient}, Asynchronous Advantage Actor-Critic (A3C)~\cite{mnih2016asynchronous}, and Deep Deterministic Policy Gradient (DDPG) algorithms ~\cite{lillicrap2019continuous}.
\end{itemize}

DRL has been applied in a wide range of fields, including: (i) \textit{Game playing}: DRL has been used to train agents that can play complex games such as Go, chess, and poker at a high level. For example, Google's AlphaGo, the first computer program to defeat a human world champion at Go, uses a form of DRL~\cite{silver2017mastering}; 
(ii) \textit{Robotics}: DRL can be used to train robots to perform tasks such as grasping objects, walking, or flying. The advantage of DRL in this context is that it allows robots to learn from interaction with the environment, rather than relying on pre-programmed instructions ~\cite{saaybi2022covy, haarnoja2018learning, doukhi2022deep}; and 
(iii) \textit{Resource management in computer systems}: DRL can be used to optimize resources in computer systems. For example, it can be used to reduce energy consumption in a data center~\cite{ran2019deepee}, optimize the scheduling of tasks in a computing cluster~\cite{mao2016resource}, or control the flow of data in a network~\cite{jay2019deep}.
In the context of serverless computing, DRL can be used to optimize function scheduling, resource auto-scaling, and cold start~\cite{qiu2022reinforcement,vahidinia2022mitigating}, for example. In this review, we opt to focus on the function scheduling and resource auto-scaling as these are complex and dynamic problems that require intelligent and adaptive solutions, which DRL is well-suited to provide.

\section{Deep Reinforcement Learning for Resource Auto-Scaling and Serverless Scheduling}
Recent studies have started to explore the application of DRL in serverless computing, particularly in the areas of function scheduling and resource scaling. 

\subsection{DRL for function scheduling in serverless computing}

Current serverless FaaS platforms typically employ basic, traditional scheduling algorithms to distribute function calls. However, they often overlook FaaS-specific traits, like swift shifts in resource use and the dynamic activation and idling of functions~\cite{yu2021faasrank}. Therefore, they present FaaSRank, a function scheduler for serverless Function-as-a-Service (FaaS) platforms. FaaSRank uses DRL to learn scheduling policies based on monitored information from servers and functions. It is implemented in Apache OpenWhisk, an open-source FaaS platform, and evaluated against other baseline schedulers using real-world serverless workload traces provided by Microsoft Azure. The results showed that FaaSRank sustained on average a lower number of inflight invocations (i.e., the number of function instances being executed) 59.62\% and 70.43\% as measured on two clusters respectively. 

Rather than solely concentrating on the cloud setting, \citet{dehury2021def} explored deploying serverless functions across both fog and cloud environments. To tackle the problem of effectively allocating functions between these environments, they introduced DeF-DReL, a DRL-driven serverless function scheduler. DeF-DReL determines the optimal split of user requests to be handled by fog and cloud, considering factors like user proximity to fog nodes, latency, user and application priorities, and resource needs. The findings indicate that the DeF-DReL agent adopts a function distribution approach that exhibits contrasting distribution patterns compared to the other two traditional scheduling algorithms examined.

\citet{yao2023performance} focus on optimizing serverless task offloading in an edge-IoT environment. The authors propose an experience-sharing deep reinforcement learning-based  (ES-DRL)  method for optimizing task offloading in serverless edge computing. ES-DRL uses a distributed learning architecture and a population-guided policy search method to enhance performance and avoid local optima. ES-DRL's main objective is to decide whether to execute a computation task locally on an IoT device or to offload it to an edge server taking into account metrics like CPU utilization and energy consumption. The method is tested against existing DRL-based task offloading methods and shows a reduction in average latency by up to approximately 17\%.

\citet{mampage2023deep} present a DRL technique based on DQN~\cite{Mnih2015} for function scheduling in multi-tenant serverless computing environments. The proposed method addresses challenges such as resource contention and the ephemeral nature of serverless functions. The DRL model optimizes the trade-off between application response time and resource usage cost. The technique, evaluated using the Kubeless serverless framework, shows significant improvements in both response time and resource usage cost compared to baseline techniques, demonstrating the effectiveness of DRL in managing resource allocation in multi-tenant  serverless environments.

\subsection{DRL for Resource Auto-Scaling}

\citet{schuler2021ai} investigate the use of RL to optimize auto-scaling in serverless computing environments. The authors argue that serverless computing, while offering significant advantages such as scalability and cost-effectiveness, presents challenges in resource management due to fluctuating demand. They propose an RL model based on Q-learning to dynamically determine the optimal level of concurrency for individual workloads in a serverless environment. The authors implemented this model in Knative, an open-source serverless platform, and evaluated its performance. The results showed that the proposed model was able to learn an effective scaling policy within a limited number of iterations, improving performance compared to the default auto-scaling configuration. While the results obtained are specific to a single application, limiting their general applicability, the authors nevertheless conclude reinforcement learning offers a promising approach for optimizing auto-scaling in serverless computing environments. 
\cite{yuvaraj2021improved} improved the performance of Q-learning by combining it with Grey Wolf Optimization (GWO) algorithm: a nature-inspired metaheuristic method that iteratively evaluates candidate solutions to approach an optimal one~\cite{mirjalili2014grey}.

\citet{zafeiropoulos2022reinforcement} studied the effectiveness of three (D)Rl agents (namely, Q-learning, DynaQ+, and DQL) to optimize auto-scaling in serverless computing. The agents are trained in real and simulated environments using the Kubeless platform. The results validate the effectiveness of these agents in balancing application performance and resource usage, demonstrating the potential of RL in optimizing serverless computing.

\citet{bensalem2023scaling} present a DRL and RL approach to auto-scaling serverless functions in edge networks. The authors argue that traditional cloud-based auto-scaling mechanisms are not directly applicable to edge networks due to their distributed nature, the complexity of optimal resource allocation, and the delay sensitivity of workloads. They propose a (D)RL-based solution to efficiently scale and allocate resources for serverless functions in edge networks. The paper compares these RL and DRL algorithms with empirical, monitoring-based heuristics, particularly for delay-sensitive applications. The simulation results show that the RL algorithm outperforms standard, monitoring-based algorithms in terms of the total delay of function requests, achieving an improvement in delay performance by up to 50\%. Results showed that RL is as good as DRL, which allows
us to use the basic RL as a fast and efficient solution.

\citet{qiu2022reinforcement,qiu2022simppo} discuss the application of DRL for resource management in multi-tenant serverless platforms. The authors present a customized multi-agent DRL algorithm based on Proximal Policy Optimization (MA-PPO). They argue that the state-of-the-art single-agent RL algorithm (S-RL) suffers from high function tail latency degradation on multi-tenant serverless FaaS platforms and struggles to converge during training. In contrast, MA-PPO allows each agent to be trained until convergence and provides online performance comparable to S-RL in single-tenant cases with less than 10\% degradation. Furthermore, MA-PPO offers a 4.4x improvement in S-RL performance in multi-tenant cases. The authors evaluate their approach using both real-world and synthetic function invocation patterns, and they use benchmarks from widely used open-source FaaS benchmark suites.

\citet{wang2019distributed} propose SIREN, a distributed machine learning framework based on serverless architecture. SIREN utilizes DRL to dynamically control the number and memory size of the serverless functions used in each training epoch. It is able to achieve higher parallelism and elasticity while reducing system configuration overhead. The prototype implementation on AWS Lambda showed a reduction in model training time by up to 44\% compared to traditional machine learning training benchmarks on AWS EC2 at the same cost. The paper argues that serverless architecture is a more cost-effective and manageable solution for machine learning practitioners and data scientists.

\section{Methodology}

In this study, we conduct a systematic review of the application of RL and DRL in serverless computing, focusing on function scheduling and resource scaling. 
We perform a comprehensive literature search in several databases, including IEEE Xplore, ACM Digital Library, Springer, and Google Scholar. The search terms include combinations of "serverless computing", "function scheduling", "resource scaling", "deep reinforcement learning", and "DRL". We also manually search the reference lists of the included studies to identify additional relevant papers.
We include studies that apply RL and/or DRL to serverless computing, particularly in the areas of function scheduling and resource scaling. We exclude studies that do not use DRL, do not focus on serverless computing, or do not provide sufficient detail about the DRL models and their performance.


\section{Findings and Discussion}

Our systematic review reveals a significant interest in the application of DRL for function scheduling and resource scaling in serverless computing. The reviewed studies demonstrate the potential of DRL to address the dynamic and complex challenges of serverless computing. It is evident that DRL can learn from the environment and adapt to changing conditions, which are key characteristics of serverless computing environments. 

DRL-based serverless schedulers such as FaaSRank \cite{yu2021faasrank}, DeF-DReL \cite{dehury2021def}, ES-DRL \cite{yao2023performance}, and DQN-based model \cite{mampage2023deep} have shown DRL algorithms are superior to classical heuristics in optimizing serverless function scheduling. 
In terms of resource auto-scaling, Q-learning based models \cite{schuler2021ai}, Grey Wolf Optimization combined with Q-learning \cite{yuvaraj2021improved}, DRL and RL based models \cite{bensalem2023scaling,zafeiropoulos2022reinforcement}, MA-PPO \cite{qiu2022reinforcement,qiu2022simppo}, and SIREN \cite{wang2019distributed} demonstrated that DRL can more optimally manage dynamic resource demands in serverless computing environments compared to the current state of the art algorithms.

Despite the promising results demonstrated by the reviewed studies, there are several challenges and open questions that need to be addressed.

First, the problem of efficient function scheduling and resource auto-scaling in serverless computing is a complex, multi-dimensional optimization problem. It requires managing trade-offs among competing objectives such as cost, performance, and resource utilization. The existing DRL models mainly focus on one or two objectives, leaving the multi-objective optimization an open problem.
Second, serverless computing operates in a highly dynamic and uncertain environment, with fluctuating workloads and resource availability. The current DRL models need to deal with this high level of uncertainty and variability. Incorporating robustness 
and uncertainty quantification in the learning process is therefore crucial for the successful application of DRL in serverless computing.
Third, the evaluation and comparison of different DRL models is challenging due to the lack of standardized benchmarks and performance metrics and datasets. Each study uses different evaluation scenarios, metrics, and baselines, making it difficult to compare the results and identify the best-performing models.
Finally, there are practical challenges in deploying DRL in serverless computing, such as the high computational cost of DRL, the difficulty in debugging and understanding the learned policies, and the sensitivity of DRL to hyperparameter settings.


\section{Future Research Directions}

Based on the findings and discussion, we suggest several directions for future research in the application of DRL for serverless computing:

\begin{itemize}
    \item \textbf{Multi-objective optimization:} Future research should investigate DRL models that can handle multiple objectives simultaneously and balance trade-offs among cost, performance, and resource utilization.
    
    \item \textbf{Robustness and uncertainty quantification:} Developing DRL models that can handle uncertainty and variability in serverless computing environments is crucial. Techniques such as adversarial training, Bayesian neural networks, ensemble methods, and meta-learning should be explored to develop robust DRL-based serverless solutions. 
    
    
    \item \textbf{Benchmarks and performance metrics:} The development of standardized benchmarks and performance metrics for evaluating and comparing DRL models in serverless computing would be beneficial. 
    
    \item \textbf{Practical deployment:} Research is needed to address the practical challenges in deploying DRL in serverless computing, such as reducing the computational cost of DRL, improving the interpretability of learned policies, and making the DRL models more robust to different hyperparameter settings. Solutions could include developing more efficient training algorithms, creating tools for visualizing and interpreting DRL policies, and designing hyperparameter tuning strategies specifically for serverless computing environments.

    \item \textbf{Real-time adaptation:} As serverless computing environments are highly dynamic, DRL models need to adapt to changes in real-time. Future work could focus on online learning strategies and methods for continual learning, enabling DRL models to adjust their policies as the environment changes.

    \item \textbf{Collaborative learning:} In the context of collaborative learning in serverless computing, federated learning emerges as a promising approach. Federated learning enables multiple (potentially untrusted) stakeholders to collaboratively train a model without directly sharing their data. Instead, they share model updates or gradients, ensuring data privacy. This decentralized training approach is particularly suitable for serverless computing environments, where data might be distributed across various nodes or locations. By leveraging federated learning, DRL models can benefit from diverse data sources without compromising on data security or privacy, leading to more robust and generalized policies.

    \item \textbf{Privacy and security:} As DRL models interact with the environment and collect data for learning, it's essential to consider privacy and security issues. Future research could explore methods for privacy-preserving learning and mechanisms to ensure the security of DRL models in serverless computing environments.
    \end{itemize}

\section{Conclusion}

In conclusion, the application of DRL in serverless computing presents promising opportunities and significant challenges. While existing studies have demonstrated the potential of DRL for function scheduling and resource scaling, much work remains to be done. Future research should focus on addressing the challenges and exploring the suggested research directions to advance the field and realize the full potential of DRL in serverless computing.

\bibliographystyle{ACM-Reference-Format}
\bibliography{main}

\end{document}